\documentclass[%
a4paper]{mye}
\usepackage[latin1]{inputenc}
\usepackage{amsmath} 
\usepackage{micce}
\usepackage{graphicx}
\usepackage{natbib,times}

\renewcommand{\title}[1]{
  \begin{center}
{\LARGE#1}\vskip 1.5em    
  \end{center}}
\renewcommand{\author}[2][Q]{\textbf{#2}\protect$^{#1}\protect$}
\renewcommand{\affil}[2][]{$^{#1}$#2\par}

\begin{document}
\thispagestyle{empty}
{\title{3-D extent of the main ionospheric trough\\---a case study}%

\flushleft

\author[1]{Mikael Hedin},
\author[1]{Ingemar Häggström},
\author[1]{Asta Pellinen-Wannberg},
\author[1]{Laila Andersson},
\author[1]{Urban Brändström},
\author[1]{Björn Gustavsson},
\author[1]{Åke Steen},
\author[2]{Assar Westman},
\author[2]{Gudmund Wannberg},
\author[3]{Tony van Eyken},
\author[4]{Takehiko Aso},
\author[5]{Cynthia Cattell},
\author[6]{Dave Klumpar}
\author[7]{and Charles W. Carlson}
\vskip 1em
\affil[1]{Swedish Institute of Space Physics, Box
    812, S-981 28 Kiruna, Sweden.}
\affil[2]{EISCAT Scientific Association,
    Box 812, S-981 28 Kiruna, Sweden.}
\affil[3]{EISCAT Scientific Association,
    Postboks 432, N-9170 Longyearbyen, Norway.}
\affil[4]{National Institute of Polar
    Research, 1-9-10 Kaga, Itabashi-ku, Tokyo 173, Japan.}
\affil[5]{School of Phys. and Astr., Univ.
    of Minnesota, MN 55455, USA.}
\affil[6]{Space Sciences Laboratory, Univ.
    of California, Berkeley, CA 94720, USA.}
\affil[7]{Lockheed-Martin Palo Alto Research Labs,  CA 94304, USA.}
\vskip 2em
}

\begin{abstract}
\noindent
  The EISCAT radar system has been used for the first time in a
  four-beam meridional mode.  The FAST satellite and ALIS imaging
  system is used in conjunction to support the radar data, which was
  used to identify a main ionospheric trough.  With this large
  latitude coverage the trough was passed in 2½ hours period.  Its
  3-dimensional structure is investigated and discussed.  It is found
  that the shape is curved along the auroral oval, and that the trough
  is wider closer to the midnight sector.  The position of the trough
  coincide rather well with various statistical models and this trough
  is found to be a typical one.
\end{abstract}

\section{Introduction}
\label{sec:introduction}

The main ionospheric trough is a typical feature of the sub-auroral
ionospheric F-region, where it is manifest as a substantial depletion
in electron concentration.  It is frequently observed in the nighttime
sector, just equatorward of the auroral zone.  This trough is often
referred to as the ``main ionospheric trough'' or ``mid-latitude
trough'' to distinguish from troughs in other locations.  The polar
edge of the trough is co-located with the auroral zone.  The
equatorward boundary is less distinct, consisting of a gradually
increasing amount of electrons, towards the plamasphere, which could
be called the normal ionosphere.  Inside the trough, there are
electric fields present, giving rise to westward ion convection.
Extensive reviews of modelling and observations of the main
ionospheric trough are given by \citet{moffett1983jatp} and
\citet{rodger1992jatp}.

The present study was performed as a part of International Auroral
Study (IAS).  The goal for IAS was to provide simultaneous
observations from ground and space of auroral processes.  An important
part of IAS was the FAST \citep{carlson1998grl} satellite.  FAST was
planned with significant ground-base support, \ie control station and
supporting scientific instrument, in Alaska, but as all polar orbiting
satellites also passes over northern Scandinavia, it is well suited
for coordination with ground-based instrument there as well.  The most
important instrument, besides the EISCAT radars, is ALIS
\citep{braendstroem1994}, the camera network in northern Sweden for
auroral imaging.

\section{The 4-beam EISCAT radar configuration}

Previous radar experiments have used different scanning patterns to
determine the topography of the trough; \citet{collis1988jatp} used a
wide latitude scan (EISCAT common program experiment CP-3),
\citet{collis1989asr} used a small 4-position scan (CP-2), and
\citet{jones1997anngeo} used a combination of a wide scan to find the
trough, and a narrow scan to observe the structure.  Also a single
position tri-static radar mode (CP-1) has been used by
\citet{haeggstroem1990jatp}.  These methods all give a trade-off
between time and space resolution---the more scan points the longer
the time before subsequent observations of the same position, and for
small and fast scans the range in space is often insufficient

Here we use for the first time all the EISCAT radars in a four-beam
configuration close to the meridian plane to get a wide area of
observation without losing time resolution.  The EISCAT Svalbard Radar
(ESR) \citep{wannberg1997rs} is field aligned (elevation 81.5°) at
invariant latitude (ILAT) 75.2°N.  The mainland tri-static UHF system
\citep{folkestad1983rs} is also field aligned (elevation 77.4°) at
ILAT 66.3°N.  Between these, the Tromsø VHF radar is used in a
``split-beam mode'', with the eastern antenna panels pointing north
(70° elevation) and the western panels pointing vertical.  The data is
split into two sets, VHF-N (north) and VHF-V (vertical).

The combined ``meta-radar'' has a huge fan-like observation area
around 70°N--80°N in geographic latitude.

\section{Characteristics of the observed trough}
\label{sec:trough-char}

\begin{figure*}[p]
  \begin{center}
    \includegraphics%
[width=\textwidth]%
{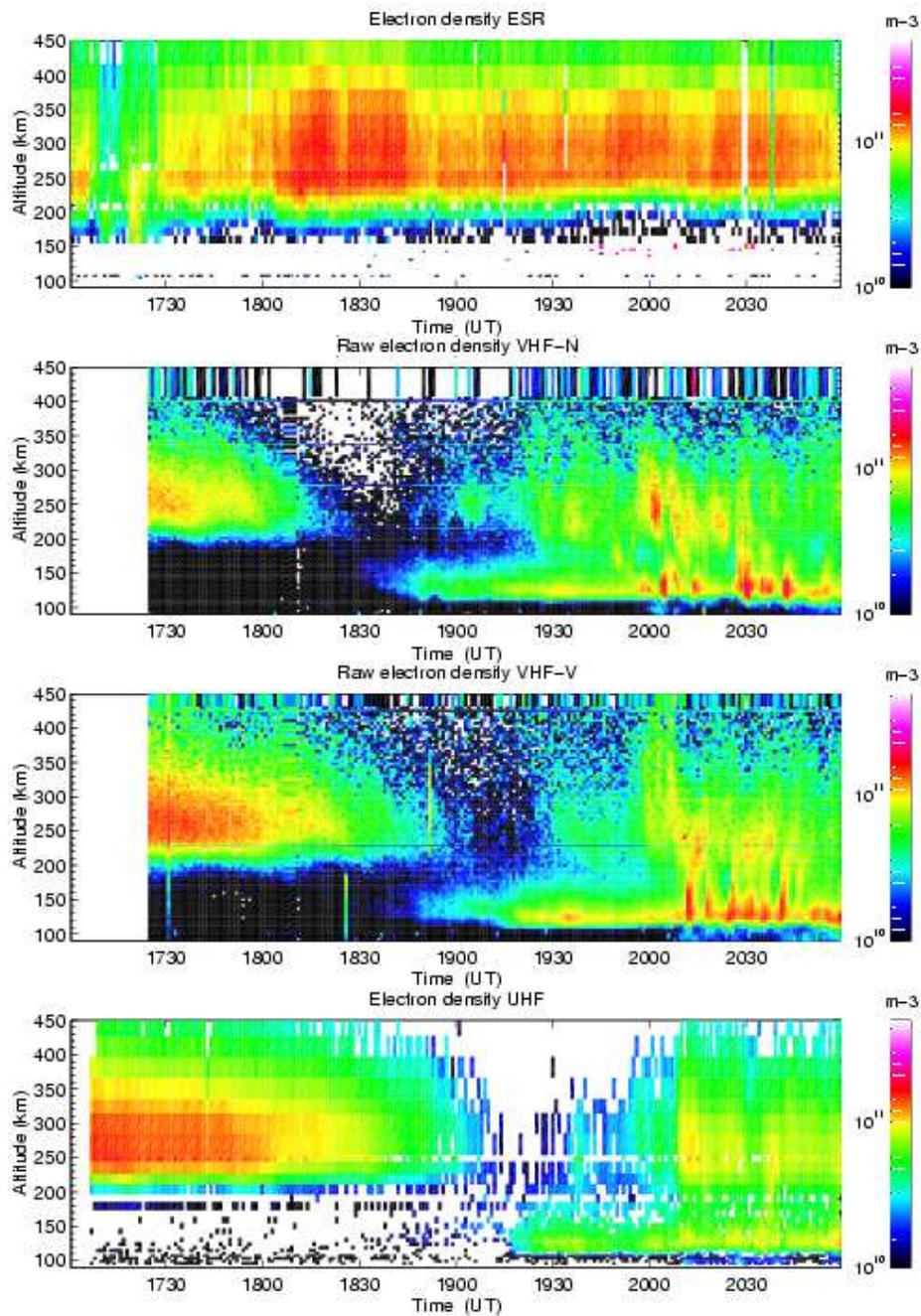}
    \caption{EISCAT electron density (ESR and UHF) and raw electron density
      (VHF-N and VHF-V) plots for 970314.  The electron concentration
      (in m$^{-3}$) is colour coded, with UT on horizontal and height
      (km) on vertical axis.    \label{fig:panel.plot}}
  \end{center}
\end{figure*}

The radar observations of the trough in question are shown in
figure~\ref{fig:panel.plot}.  The trough is seen as the clear decrease
in electron density, and we determine the time in UT for the radars
passing under the trough to be 1710--1720, 1820--1900, 1845--1925 and
1900--1940 respectively.  The more prominent density increase for the
Tromsø sites after 2000 UT is not the edge of the trough, rather
typical F-region blobs, poleward of the trough.

In figure~\ref{fig:panel.plot} we see that the trough extends through
all of the F-region, but not in the E-region.  Note that white color
is both highest density and no usable data, but generally no data is
due to low density and this is seen to be the case here in the trough.
In the northern part of the trough (later time in the data), we can
actually see some typical weak ionization in the E-region, interpreted
as diffuse aurora.

\begin{figure}
  \begin{center}
    \includegraphics{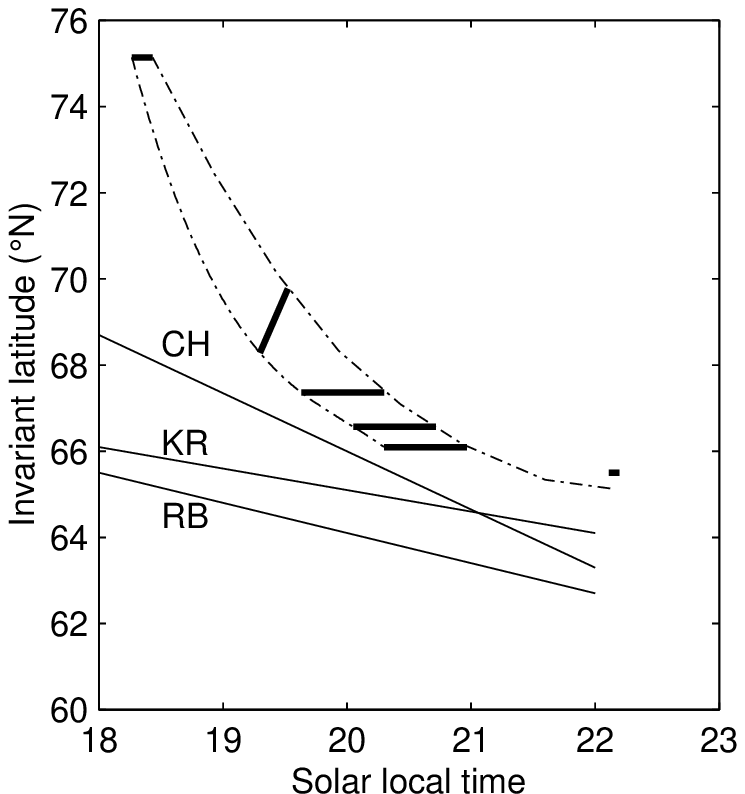}
    \caption{The thick lines show the coordinates in ILAT and SLT
      (solar local time) of the actual pass of the trough as
      determined from (high to low latitude) ESR, FAST, VHF-N, VHF-V
      and UHF.  The dash-dotted line is an estimate of the trough
      boundaries, extrapolated to later time from these direct
      measurements.  The diffuse aurora observed with ALIS is
      indicated by the small spot in the latest part of the plot, just
      above the border of the estimated trough boundary.  The solid
      lines show predictions for location of trough minima from models
      discussed in the text.    \label{fig:troug-pos}}
  \end{center}
\end{figure}

We compare the actual location of the trough minimum with predicted
positions from models by
\citet{collis1988jatp,koehnlein1977pss,rycroft1970jgr}
respectively\footnote{with \citet{rycroft1970jgr} changed to invariant
  latitude as suggested by \citet{koehnlein1977pss}}---all linear in
SLT (solar local time) and $\mathit{Kp}$.  For the present day,
$\mathit{Kp}$ values are $1-$ before and 1o after 1800 UT.  In
figure~\ref{fig:troug-pos}, these model values are shown together with
the actual position of the whole trough as determined from ESR, FAST,
VHF-N, VHF-V and UHF respectively (from high to low latitude).  We see
that the deviations from the model predictions are substantial, and
conclude that the linear models are not adequate for use over a wide
range in latitude and time.  This is not surprising because the fits
used to construct the equations all had a big spread, even though the
correlations were quite good.

A more recent study based on satellite data is made by
\citet{karpachev1996asr}, where they use both latitude and longitude
to make a statistical model not restricted to linear relations.  In
this model, the time used is magnetic local time (MLT), which is
reasonable because the trough structure is governed by the earth
magnetic field.  They use both linear and non-linear time dependency,
and find that for midnight hours, the difference is rather small,
compared to the large spread in the data, but favours the non-linear
time dependency.

From the start time of when each radar beam enters the trough, as seen
in figure~\ref{fig:panel.plot}, we can estimate the trough apparent
southward speed to be around 12 km/min from Longyearbyen to Tromsø,
and around 4 km/min between the Tromsø beams.  Assuming that the
trough appears to pass over ESR and between the mainland radars with
the respective speed, we can make a crude estimate of the vertical
shape of the trough and the width: From the plot in
figure~\ref{fig:panel.plot}, we estimate that the southward (early)
wall, taken as the border of blue and green color, has an inclination
of 35° for the VHF-N beam and 21° for VHF-V.  This should be compared
with the direction of the magnetic field at Tromsø, zenith angle 13°
southwards.  If we assume the trough wall to be field aligned, the
anticipated inclination is 33° (20+13) for VHF-N and 13° for VHF-V, in
good agreement with the rough estimate.  In the plot of the field
aligned beams, ESR and UHF, the trough wall appears vertical, which
means it is field-aligned.  The passing time for Longyearbyen is 10
minutes, which, using the above speeds, gives a width of 120 km, and
for the Tromsø beams the time is 40 minutes, which gives a width of
160 km, both in north-south direction.  This assumes the trough to
consist of two straight parts, one extending from over Longyearbyen
and Tromsø, the other over the Tromsø beams, instead of the actual
arc-like shape as seen in figure~\ref{fig:troug-pos}, but it will
anyway give an indication of the size.  The east-west size can be
estimated by using the earth rotation, this gives a velocity of 6.1
km/min, and a size of 61 km for Longyearbyen and around 10 km/min with
size 400 km over Tromsø.  The actual width, measured perpendicular
over the trough extent is then 54 km for the part passing over
Longyearbyen and 149 km for the part passing over Tromsø.  That is,
the trough radial width is actually broader equatorwards, closer to
magnetic midnight.  This is shown in figure~\ref{fig:troug-pos}.

The FAST satellite does not carry sounding instruments, so trough
signatures are not as obvious to detect.  The signature used is the
precipitation north of the trough and the electric field associated
with the ion ($E\times B$) drift (absent south of the trough).  It is
known from earlier studies \citep{collis1988jatp} that there is a
westward ion convection in the trough, which is absent outside.  The
satellite passes over the trough 1742--1743 UT as shown in
figure~\ref{fig:troug-pos}.

During the night, the auroral imaging system ALIS was not operating
continuously because there was no significant aurora and it was partly
cloudy. However, some pictures were taken at relevant times, and they
show faint diffuse aurora around the time when ALIS passes under the
trough poleward boundary.  If we plot the position of diffuse aurora
from ALIS, it located just in the northern part of the trough region
as extrapolated from the other more direct measurements in
figure~\ref{fig:troug-pos}.

If we compare the observed trough with the relevant typical features
of the mid-latitude trough as described in \citet{moffett1983jatp}, we
can say that the present trough is quite typical.

\section{Conclusions}
\label{sec:conclusions}

The EISCAT facility has been used in a new four-beam meridional mode.
The ESR and UHF system are field-aligned, and the VHF system is split
in two beams in between.  This gives a very wide latitude range for
observation, well suited to study ionospheric structures moving over
large range in short time.  With this configuration, no time
resolution is lost.

In this case, the main ionospheric trough has been observed as high as
75°N ILAT down to 66°N ILAT.  The observed trough is a quite typical
one--it has all the common features known from earlier studies.  The
present observation is made when the $\mathit{Kp}$ index was low and
stable, and so the trough was also quite non-dramatic, but this means
that the trough is rather stationary over the big area of observation.
This rules out significant time variations of the trough position, as
would have been the case in a more active environment, enabling
investigation over a relatively long period as the earth moves under
it.

If we calculate the apparent southward motion of the trough, the speed
is over 10~km/min between Longyearbyen and Tromsø, but about 4~km/min
between the Tromsø sites.  This is consistent if the trough is really
an oval shape, so that Longyearbyen passes under it close to
perpendicular, but for Tromsø latitudes the direction of passage is
much more oblique, thereby the apparent southward motion is much
slower.  The trough is also seen to be wider towards magnetic
midnight.  The earlier proposed linear equations for trough motions
are shown not to be valid over the latitude range in question.

Coordinated studies poses substantial difficulties, most of which are
not scientific but rather administrative or probabilistic by nature.
Nevertheless, if one measurement (EISCAT in this case) is good, the
others can often be used to extract some extra information in support.

\begin{acknowledgement}
  We gratefully acknowledge assistance of the EISCAT staff. The EISCAT
  Scientific Association is supported by France (CNRS), Germany (MPG),
  United Kingdom (PPARC), Norway (NFR),Sweden (NFR), Finland (SA) and
  Japan(NIPR).
\end{acknowledgement}

\bibliography{strings,a_z}

\bibliographystyle{egs}

\end{document}